\documentclass{article}
\usepackage[margin=1.5in]{geometry}
\usepackage{graphicx}
\usepackage{authblk}
\usepackage{amsmath}
\usepackage[font=small,width=4.25in,labelsep=period]{caption}
\usepackage[colorlinks=true,urlcolor=urlblue,linkcolor=black,citecolor=black,breaklinks=true,bookmarks=false]{hyperref}
\newenvironment{backmatter}{%
  \small%
  \newcommand{\bmsection}[1]{\par\medskip\noindent{\bfseries ##1.\enspace}}%
}{}
\renewcommand\figurename{Fig.}
\newcommand\norm[1]{\left\lVert \mathbf{#1} \right\rVert}

\title{Phase Diverse Phase Retrieval for Microscopy: Comparison of Gaussian and Poisson Approaches}
\author[1]{Nikolaj Reiser\thanks{nikolaj.reiser@gmail.com}}
\author[2]{Min Guo}
\author[2]{Hari Shroff}
\author[1]{Patrick J. La Riviere}
\affil[1]{\small University of Chicago, Department of Radiology, Chicago, Illinois, USA}
\affil[2]{Laboratory of High Resolution Optical Imaging, National Institute of Biomedical Imaging and Bioengineering, National Institutes of Health, Bethesda, Maryland, USA}

\begin{document}
\definecolor{urlblue}{RGB}{46,46,177}

\maketitle

\begin{abstract}
Phase diversity is a widefield aberration correction method that uses multiple images to estimate the phase aberration at the pupil plane of an imaging system by solving an optimization problem. This estimated aberration can then be used to deconvolve the aberrated image or to reacquire it with aberration corrections applied to a deformable mirror. The optimization problem for aberration estimation has been formulated for both Gaussian and Poisson noise models but the Poisson model has never been studied in microscopy nor compared with the Gaussian model. Here, the Gaussian- and Poisson-based estimation algorithms are implemented and compared for widefield microscopy in simulation. The Poisson algorithm is found to match or outperform the Gaussian algorithm in a variety of situations, and converges in a similar or decreased amount of time. The Gaussian algorithm does perform better in low-light regimes when image noise is dominated by additive Gaussian noise. The Poisson algorithm is also found to be more robust to the effects of spatially variant aberration and phase noise. Finally, the relative advantages of  re-acquisition with aberration correction and deconvolution with aberrated point spread functions are compared. 
\end{abstract}

\section{Introduction}

Images acquired in microscopes and other imaging systems are often degraded by phase aberrations. Phase aberrations occur when the spherical wavefronts emitted from a sample become distorted by misaligned optics or refractive index (RI) mismatches and variations \cite{Goodman2005}. These aberrations cause the spherical wavefronts to converge into spots much larger than the diffraction limit on the imaging sensor, decreasing the sharpness and contrast of images.

In order to restore diffraction-limited performance of optical systems, aberrations need to be corrected, either using hardware like a deformable mirror (DM) or with software deconvolution. For this to be achieved, the aberrations must first be detected and measured, so they can be properly compensated.  Phase aberrations can be detected with hardware-based methods, like a Shack-Hartmann wavefront sensor (SHWFS), or with software-based methods. Software-based methods include optimizing a predefined image metric over DM configurations and phase retrieval (PR) \cite{Booth2014} \cite{Ji2017}.

In this paper, phase diverse phase retrieval, also known as phase diversity (PD) will be used to estimate aberrations occurring in widefield images. In PD, multiple images are acquired, each with a different but known phase aberration purposefully introduced. With this set of images, the unknown phase aberration can be estimated by optimizing an objective function appropriate for the assumed image noise model. This technique was first derived for extended objects by Gonzalvez \cite{Gonsalves1979} \cite{Gonsalves1982} in the case of a single additional diversity image assuming Gaussian noise. Paxman et al. later generalized the derivations to multiple images with both Gaussian and Poisson noise models, but presented no results or implementation \cite{Paxman1992}.

There have been very few studies of PD in microscopy and all have used the Gaussian approach \cite{Turaga2010} \cite{Kner2013} \cite{Zhang2017}. Here, the Poisson-based approach proposed by Paxman is implemented and compared to the Gaussian one. The Poisson estimator may offer improved aberration estimation over the Gaussian likelihood function according to Cramer-Rao lower bound estimates for extended objects \cite{Dolne2005}. PD with the Poisson estimator has had some limited use in astronomy in deconvolution \cite{Schulz1997} \cite{Archer2013} and wavefront estimation \cite{Luna2005} \cite{Seldin1994}, but with little evaluation of aberration estimation performance and no comparison to the Gaussian estimator.

\section{Theory}

\subsection{Image Formation}
In incoherent fluorescence microscopy, the noiseless image formation process within the scalar diffraction approximation is modeled as 
\begin{align}
&g(\mathbf{x}) = s(\mathbf{x}) * f(\mathbf{x}) \\
&s(\mathbf{x}) = \mid \!\! \mathcal{F} (H(\mathbf{ u }))  \!\! \mid ^2 \,, \label{PSF}
\end{align}
where $g$ is the image, $s$ is the point spread function (PSF), $f$ is the object, and $H$ is the generalized pupil function (GPF). Coordinates $\mathbf{x}$ and $\mathbf{u}$ are 2d spatial coordinates in the image/object plane and pupil plane respectively. $\mathcal{F}$ denotes a Fourier transform and $*$ denotes convolution. Phase aberrations $\phi$ are described in the GPF as
\begin{equation}
H(\mathbf{u}) = P(\mathbf{u}) \exp \left( {i \, \phi \! \left( \mathbf{u} \right)} \right) \,, \label{GPF}
\end{equation}
where $P$ is the pupil function. No apodization is used:
\begin{equation} \label{pupil}
P(\mathbf{u}) = \begin{cases}
1 & \norm{u} \leq \mathrm{NA}/\lambda \\
0 & \norm{u} > \mathrm{NA} / \lambda \,.
\end{cases}
\end{equation}

To reduce the dimensionality of the problem,  the phase aberrations are expressed using a basis of Zernike polynomials:
\begin{equation}
\phi(\mathbf{u}) = \sum_j c_j Z_j \left( \frac{\lambda}{\mathrm{NA}} \mathbf{u} \right) \label{zern} \,.
\end{equation}
Zernike polynomials have both a double-index representation ($Z^m_n$) and a single-index representation ($Z_j$). We use a single-index representation following the Noll convention \cite{Noll1976}. The factor of $\frac{\lambda}{\mathrm{NA}}$ is to align the domain over which the Zernike polynomials are orthogonal with the nonzero domain of the pupil function $P$.

\subsection{Phase Diversity}
PD requires multiple images with known phase aberrations (referred to as diversity phases and diversity images) that are purposefully introduced:
\begin{align} \label{phase_diversity_equations}
&g_k(\mathbf{x}) = s_k(\mathbf{x}) * f(\mathbf{x}) \\
&s_k(\mathbf{x}) = \mid \!\! \mathcal{F} \left[ H_k(\mathbf{u})\right]  \!\! \mid ^2 \\
&H_k(\mathbf{u}) = H(\mathbf{u}) \exp i \, \theta_k (\mathbf{u})  \,,
\end{align}
where $\theta_k$ are the diversity phases. While there are no restrictions on diversity phase selection, a `good' choice of diversity phase is essential for creating a likelihood function that leads to an accurate phase estimation. For convenience,  we only consider defocus for the diversity phase, as it is rotationally symmetric and easily introduced in most optical systems by moving the sample relative to the lens, which allows for use of PD even in systems without a DM.
Note that the term defocus is used in the literature to refer to both Zernike polynomial $Z^0_2$ and to the defocus function \cite{Kner2013}:
\begin{align}
&\theta_k  \! (\mathbf{u}) = z_k \gamma({\mathbf{u}}) \\
&\gamma(\mathbf{u}) = \frac{2 \pi n}{\lambda}\sqrt{1- \left( \frac{\lambda \norm{u} }{n} \right)^2} \,,
\end{align}
where $\gamma$ is the defocus function, $n$ is the RI of the medium, and $z_k$ is the distance from the focal plane for the $k^{\mathrm{th}}$ diversity phase. The Zernike polynomial defocus is an approximation of the defocus function as described in \cite{HieuThao2020}.  Here, we use the defocus function, not its Zernike approximation, to model the use of defocus as a diversity phase.

Using three diversity phases, with $z_0$ set to $3 \lambda$, $z_1$ set to $0 \lambda$, $z_2$ set to $-3 \lambda$, gave relatively accurate estimations in a wide range of conditions. On a case by case basis, a more optimal set of diversity phases can be chosen, but for simplicity, this diversity phase scheme is used throughout the paper.

\section{Phase Estimation}
To estimate the unknown aberration $\phi$,  maximum-likelihood expectation maximization (MLEM) is used. Paxman et al. derive likelihood functions (LF) in \cite{Paxman1992} for additive Gaussian noise and Poisson noise. In the present paper, these LF's are both maximized and compared in a variety of situations. Besides the noise type, the LF's differ significantly in implementation. In the Gaussian LF, the  maximal likelihood object has a closed-form solution in terms of the acquired images and aberration parameters. The Gaussian LF still estimates the object, just not explicitly. In the Poisson LF, the object needs to be explicitly estimated along with the aberration parameters. 

\subsection{Gaussian}
The Gaussian LF to be maximized is derived in \cite{Paxman1992}, equation 13:
\begin{equation}
\mathcal{L}(\vec{c}) = - \sum_k \sum_\mathbf{x} \left[ d_k(\mathbf{x}) - f (\mathbf{x}) *s_k(\vec{c}, \mathbf{x}) \right]^2 \,,
\end{equation}
where $d_k$ is the measured noisy image and the $\vec{c}$ in $s_k$ is used to indicate that the quantity depends on the aberration parameters. The LF is maximized using the strategy of Vogel et al. in \cite{Vogel1998}, which develops a Gauss-Newton approach and calculates both the gradient and the pseudo-Hessian of the Gaussian likelihood function. Details of our implementation are given in our supplement.

\subsection{Poisson}
The Poisson LF to be maximized is also derived in \cite{Paxman1992}, equation 29:
\begin{equation}
\mathcal{L}(\vec{c}) = \sum_k \sum_\mathbf{x} \left[ d_k(\mathbf{x}) \, \mathrm{ln} \, g_k(\vec{c}, \mathbf{x}) - g_k(\vec{c}, \mathbf{x})  \right].
\end{equation}
This LF is maximized using the methods from \cite{Paxman1992}, which involve alternating updates of the aberration coefficients and the object.

The algorithm is initialized with a uniform object estimate and a small non-zero phase estimate (e.g. $10^{-10}$ for each Zernike amplitude). To begin, the aberration coefficients are updated by performing a line search over the gradient of the likelihood with respect to each Zernike coefficient. This gradient, specified in equation 42 from \cite{Paxman1992}, is given by:
\begin{equation}\label{p1}
\frac{\partial \mathcal{L}(\vec{c})}{\partial c_j} = -2 \sum_\mathbf{u} Z_j \left( \frac{\lambda}{\mathrm{NA}} \mathbf{u} \right) \mathrm{Im} \left[ \sum_k H_k(\vec{c}, \mathbf{u}) \mathcal{F}^{-1} \left( h^*_k(\vec{c}, \mathbf{x}) \left( \tilde{f}(\mathbf{x}) * \frac{d_k(\mathbf{x})}{f(\mathbf{x})*s_k(\vec{c}, \mathbf{x})} \right) \right) \right] \,.
\end{equation}
The tilde denotes flipping:

\begin{equation}
\tilde{f}(\mathbf{x}) = f(-\mathbf{x}) \,.
\end{equation}
Next, equation 62 from \cite{Paxman1992} is used to update the object estimate:

\begin{equation}\label{p2}
f^{r+1}(\mathbf{x}) = f^r(\mathbf{x}) \left[ \frac{1}{\sum_k \sum_\mathbf{x} s_k(\vec{c}, \mathbf{x})} \sum_k \tilde{s}_k(\vec{c}, \mathbf{x}) * \frac{d_k(\mathbf{x})}{f(\mathbf{x})*s_k(\vec{c}, \mathbf{x})} \right] \,.
\end{equation}
These two steps are repeated until stopping conditions are met.  Using specifically \eqref{p1} and \eqref{p2} is essential for fast convergence of the Poisson LF. We used a threshold on the norm of the gradient ($\frac{\partial \mathcal{L}(\vec{c})}{\partial c_j}$) for the stopping condition.

\subsection{Other Algorithm Details} \label{other}
In order to improve algorithm runtime, images are downscaled by a factor of two. Performing this downscaling gives a substantial runtime improvement, with minimal loss of estimation accuracy for both algorithms in most situations (see section \ref{imsize_section} and \figurename{\ref{fig:imsize}}). Note that this technique only works if the image sensor is sampling at the Nyquist rate or higher.

Regularization was found to have no significant impact on phase estimation with either algorithm for both the object estimate and the aberration estimate. For the Gaussian algorithm, a small non-zero value (e.g. $10^{-10}$) was used for regularizing the object estimation. For the Poisson algorithm, the object estimation was normalized after each iteration.

\section{Image Simulation}
\begin{figure}[ht!]
\centering\includegraphics[width=13cm]{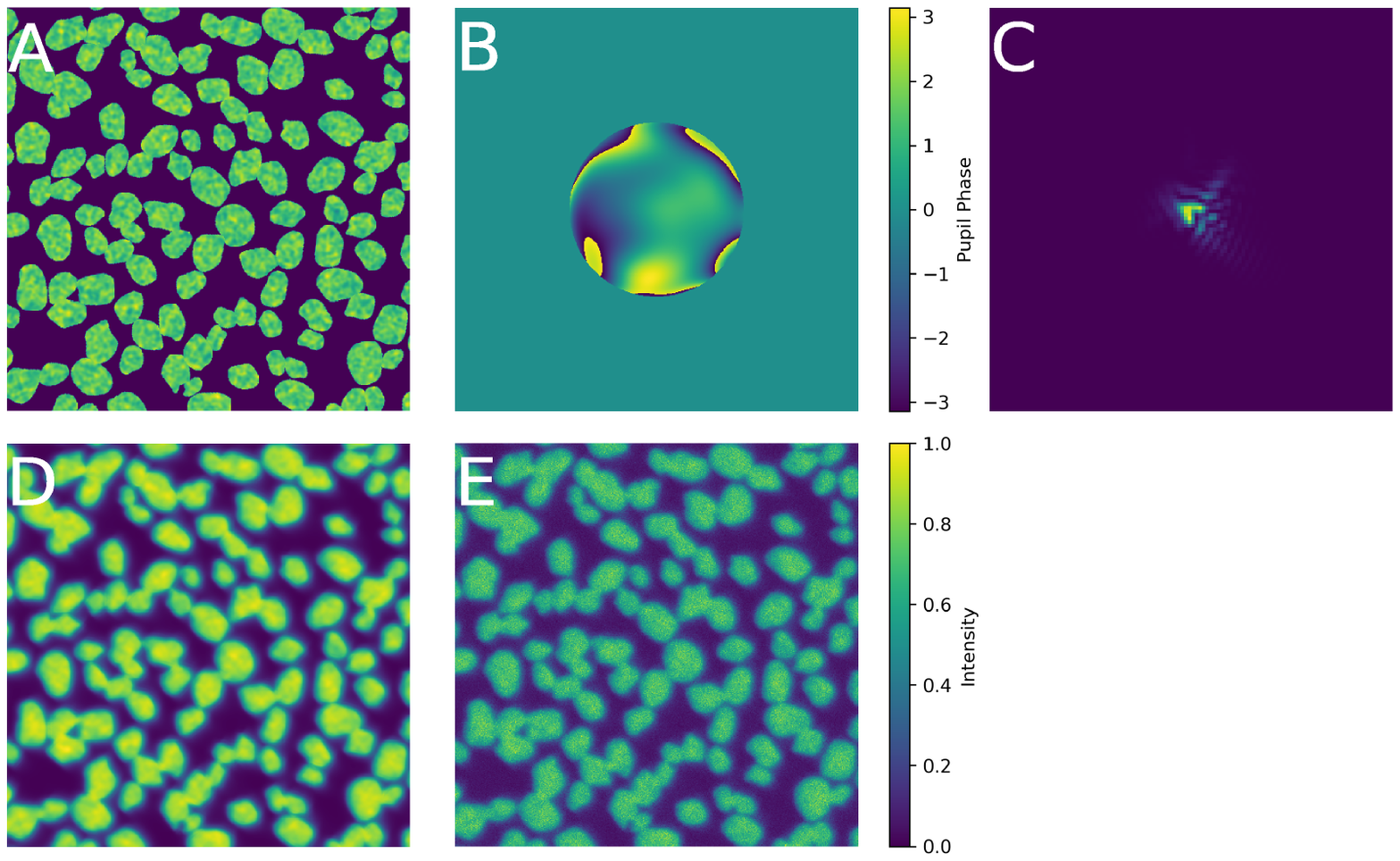}
\caption{Simulation process and corresponding images. Step 1: Generate object (\textbf{A}). Step 2: Generate phase aberration (\textbf{B}) and corresponding PSF (\textbf{C}). Step 3: Convolve PSF with object (\textbf{D}). Step 4: Simulate noise processes (\textbf{E}). Final image is normalized so that the minimum value is 0 and maximum value is 1.}
\label{fig:simulation}
\end{figure}

The process used to simulate an aberrated image is shown in \figurename{\ref{fig:simulation}}. It involves (1) creating a synthetic object, (2) generating a phase aberration of specified magnitude and its corresponding PSF, (3) convolving the PSF with the synthetic object, and (4) applying noise. These steps are introduced in greater detail in the following sections.

\subsection{Object Simulation}
\begin{figure}[ht!]
\centering\includegraphics[width=10cm]{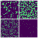}
\caption{Synthetic objects used to test PD algorithms.}
\label{fig:objects}
\end{figure}

Objects are approximated as 2D distributions and are created similar to the methods described in \cite{Lehmussola2007}. Using a synthetic object allows extensive control over the object properties (such as size of features, number of cells, cell texture, etc.). The object properties can have strong influence on the estimation accuracy of the PD algorithms, and fine control over those properties allows rapid testing of the algorithms in a variety of situations. The four simulated objects used to test the PD algorithms are shown in \figurename{\ref{fig:objects}}.

\subsection{Aberration}
To set the amount of each Zernike component present in the simulated wavefronts, uniform random values between 1 and -1 are selected for the coefficients. 42 coefficients are used, corresponding to Zernike polynomials of order 2 through 7 (j = 4 through j = 45). Polynomials of orders lower than 2 are not considered here, as they do not cause image blurring and cause no change in the LFs. 

The first 12 coefficients (n = 2 through n = 4) are scaled to a specific aberration magnitude, which is specified in units of wavelength. Aberration magnitude (also called wavefront RMS) is defined as:
\begin{align}
\mathrm{WRMS}(\phi) &= \frac{1}{2 \pi} \sqrt{ \int \phi (\mathbf{u})^2 d\mathbf{u}} \\
&= \frac{1}{2 \pi} \sqrt{ \sum_j \frac{ c_j^2 }{A_j} } \,,
\end{align}
where $A_j$ is the normalization factor $\int Z_j \left( \frac{\lambda}{\mathrm{NA}} \mathbf{u} \right) ^2 d\mathbf{u}$.

The rest of the coefficients are scaled to half of that aberration magnitude, as they are usually present in lower amplitude \cite{Azucena2010} \cite{Tao2012} \cite{Schwertner2004}. Not all coefficients need to be estimated; the large number of coefficients for image simulation is to verify that lower-order coefficients can be estimated in the presence of higher-order coefficients. Unless otherwise indicated, every simulated aberration had an aberration magnitude of $2\lambda$.

\subsection{Convolution}
\begin{figure}[ht!]
\centering\includegraphics[width=13cm]{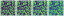}
\caption{Images shown have wavefront magnitudes of (from left to right): 0$\lambda$, 1.2$\lambda$, 2.4$\lambda$, and 4$\lambda$. Color range for each image corresponds to individual image's data range.}
\label{fig:abmag_images}
\end{figure}

After selecting the coefficients, the phase of the aberrated wavefront is calculated using \eqref{zern}, along with the GPF and PSF using \eqref{GPF} and \eqref{PSF} respectively. Next, a pristine object image is convolved with the PSF. When standard Fourier-based convolution is performed, the output image has unrealistic artifacts near the edges from the cyclic convolution. The convolution process used here still uses Fourier transforms, but the output images are cropped to more accurately simulate the infinite extent of the PSF and object. See the supplement for more details on the convolution process. Examples of noiseless, convolved images at four different aberration magnitudes are shown in \figurename{\ref{fig:abmag_images}}.

\subsection{Noise} \label{noise_sec}
\begin{figure}[ht!]
\centering\includegraphics[width=13cm]{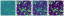}
\caption{Images shown are samples of (from left to right): high additive noise with maximum signal at 10 photons/pixel (p/px) average, high additive noise at 200 p/px, low additive noise at 10 p/px, and low additive noise at 200 p/px. Low additive noise is $I_{dc} = 1$, $\sigma_r = 2$ and high additive noise is $I_{dc} = 100$, $\sigma_r = 20$. Color range for each image corresponds to individual image's data range.}
\label{fig:noise_images}
\end{figure}

To simulate the noisy image, three noise processes are used, as in \cite{Svoboda2009}:

\begin{equation}
d_k = \mathrm{QE} \times \mathcal{P} (I^k_{ph}) + \mathcal{P}(I_{dc}) + \mathcal{N}(0,\sigma_r^2),
\end{equation}
where QE (which has a value of 0.6 in all simulations) is the quantum efficiency of the detector, $\mathcal{P}(\lambda)$ is a Poisson distributed random variable with mean $\lambda$, and $\mathcal{N}(\mu, \sigma^2)$ is a Gaussian distributed random variable with mean $\mu$ and variance $\sigma^2$. $I^k_{ph}$ represents the expected number of photons at each pixel from the convolved object and $k^\mathrm{th}$ PSF. To specify the number of expected photons per pixel (p/px, i.e. the amount of shot noise), the noiseless convolved image from the convolution step is divided by its mean, and subsequently multiplied by the desired p/px. $I_{dc}$ represents the expected number of electrons from an additive Poisson distributed noise process, such as dark noise. $\sigma_r^2$ represents the variance of the number of electrons from an additive Gaussian distributed noise process, such as read noise. $I_{dc}$ and $\sigma_r^2$ are uniform at each pixel. No ADC noise is simulated. Examples of images with different levels and combinations of noise are shown in \figurename{\ref{fig:noise_images}}. Unless otherwise indicated, every image was simulated with an average of 500 p/px and low additive noise ($I_{dc} = 1$, $\sigma_r = 2$).

\section{Simulation Results}
First, the performance of the phase estimation algorithms will be analyzed with a variety of simulation settings. Estimation accuracy of the algorithms is quantified by the residual wavefront error (RWE), which is the RMSE of the residual wavefront:
\begin{equation}
\mathrm{RWE}(\phi) = \frac{1}{2 \pi} \sqrt{ \sum_j \frac{ (\Tilde{c}_j - c_j)^2 }{A_j} } \,,
\end{equation}
where $\Tilde{c}_j$ are the ground truth Zernike coefficients.

All parameter sweeps are performed by simulating 100 image sets at each point in the sweep with randomly chosen aberrations. Error bars show standard error.

\subsection{Aberration Magnitude}
\begin{figure}[ht!]
\centering\includegraphics[width=13cm]{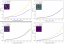}
\caption{Comparison of estimation performance over different levels of aberration and different objects. Inset indicates which object is used in each sweep.}
\label{fig:abmags}
\end{figure}

First, the accuracy of estimation is evaluated at different levels of aberration (\figurename{\ref{fig:abmags}}). Each plot shows the RWE as a function of initial aberration magnitude (wavefront RMSE). The Poisson estimation algorithm matches or outperforms the Gaussian algorithm in most objects and aberration levels.

\subsection{Image Noise}
\begin{figure}[ht!]
\centering\includegraphics[width=13cm]{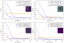}
\caption{Comparison of estimation performance over different photon levels and different objects. Inset indicates which object is used in each sweep. Low additive noise is $I_{dc} = 1$, $\sigma_r = 2$ and high additive noise is $I_{dc} = 100$, $\sigma_r = 20$.}
\label{fig:noise}
\end{figure}

Next, the algorithms are tested with different amounts of noise and different objects (\figurename{\ref{fig:noise}}). The p/px are varied, and two different additive noise strengths are used. Note that each image is simulated with the same total number of photons in the noise process, causing the sparser objects to have higher peak p/px. In the high-count regime, the Poisson algorithm matches or outperforms the Gaussian model. At lower count levels, when additive Gaussian noise dominates, the Gaussian model generally performs better. 

\subsection{Image Size} \label{imsize_section}
\begin{figure}[ht!]
\centering\includegraphics[width=13cm]{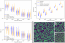}
\caption{Left: RWE and runtime vs image size. Top row: Image size was changed only with a crop. Bottom Row: Image size was changed with a crop and magnification. Downscaled indicates that the downscaling technique mentioned in section \ref{other} was used prior to running the algorithms. The short horizontal lines in the bars indicate maximum, mean, and minimum. Bar thickness indicates density of values vs RWE. Images are the largest object (\textbf{A}), the cropped image (\textbf{B}), and the cropped and magnified image (\textbf{C}).}
\label{fig:imsize}
\end{figure}

Image size variation is explored in two cases: cropping at constant magnification (\figurename{\ref{fig:imsize}}, upper left), and cropping with variable magnification (\figurename{\ref{fig:imsize}}, lower left). The first case reduces the amount of object in the FOV, but keeps it at the same resolution, while the second case keeps the same amount of object in the FOV, but at a lower resolution.

Regardless of estimation algorithm, both size-changing paradigms lead to worsening aberration estimates for smaller images. Runtime of the estimation algorithms are shown in (\figurename{\ref{fig:imsize}}, upper right), with the algorithms being run on a desktop PC with an Intel Core i7-7700K. One additional benefit to increased image size (not explored here) is that if the FOV is large enough to sample the entire object, the cyclic convolution model assumed by both estimation algorithms is accurate, giving a better aberration estimation.

With smaller scaled images, the Gaussian outperforms the Poisson algorithms. In all other cases, the Poisson matches or outperforms the Gaussian algorithms. Downscaling the images prior to running the algorithms sacrifices very little estimation accuracy, while substantially decreasing algorithm runtime.

\subsection{Spatial Variance and Phase Noise}
\begin{figure}[ht!]
\centering\includegraphics[width=13cm]{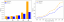}
\caption{Estimation accuracy when images are degraded by spatially varying wavefronts (left) and phase noise (right). Both simulations used lower left object (see \figurename{\ref{fig:objects}}). Bar labels in SV plot are: No Spatial Variance, Low Frequency / Low Magnitude, High Frequency / Low Magnitude, Low Frequency / High Magnitude, and High Frequency / High Magnitude.}
\label{fig:phase_noise}
\end{figure}

Biological samples have varying RIs, leading to interfaces where light is refracted, causing aberrations. However, the RI variations are different across the sample, leading to aberrations that vary spatially. The PD techniques here do not account for spatial variation, but can still find an overall `mean' aberration (\figurename{\ref{fig:phase_noise}}, left) that can be corrected over the entire image with a single DM at the pupil plane.

To simulate spatially varying aberration, the isoplanatic wavefront has a random component added to it at each point in the image. Over the entire image, these random components have zero mean, so the isoplanatic wavefront remains unchanged. The magnitude of the spatial variance (SV) was varied by scaling the random components. Images were simulated with random wavefront components that varied slowly over the image (low frequency SV) and rapidly over the image (high frequency SV). Due to the long times required to simulate images with spatially varying wavefronts, only a single aberration pattern is evaluated, albeit under these four different combinations of frequency and magnitude. The results show that both algorithms perform well in the face of most of the variation studied, although at high magnitude and high frequency the Poisson approach significantly outperforms the Gaussian approach. 

Also, some additional unknown aberration may be generated when adjusting the imaging system (with either a DM or by moving the objective lens) for acquiring the diversity images. This aberration, called phase noise, is unique to each diversity image. The mean phase noise between all the diversity images simply combines with the main aberration $\phi$ being estimated and does not violate any assumptions of the image formation. However, the residual phase noise (with zero mean) is not modelled by the image formation process, and degrades accuracy of the estimation. To measure the estimation degradation of phase noise, random vectors of Zernike components (with zero mean) are added to each individual wavefront before running the estimation algorithms (\figurename{\ref{fig:phase_noise}}, right). See the supplement for more details on SV and phase noise.

\subsection{Deconvolution vs Re-acquisition}
Having estimated the phase aberration, it can be used either to generate a PSF for deconvolution of the original image or to shape a deformable mirror for re-acquisition of a less aberrated image. To compare these two approaches, structural similarity (SSIM) \cite{Simoncelli2004} is used as an image-quality metric. The ground truth image in the SSIM comparison is an image with blur from only the diffraction process and high-order aberrations. This represents a perfectly AO-corrected image.

Deconvolution was performed with Richardson-Lucy deconvolution using all the diversity images, similar to the approach in \cite{Ingaramo2014}. Images are slightly cropped before measuring the SSIM to remove ringing artifacts (commonly occurring in Richardson-Lucy deconvolution) from comparison. Re-acqusition with a DM was simulated by applying the estimated low-order aberrations as correction phases in the pupil plane, then re-simulating the image. 

\begin{figure}[ht!]
\centering\includegraphics[width=13cm]{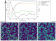}
\caption{Comparison of image quality using different correction techniques. \textbf{A}: Plot showing SSIM of various correction approaches and RWE of wavefront estimations (using Poisson LF) vs noise level. Bottom row: example of AO corrected image (\textbf{B}), deconvolved image (\textbf{C}), and uncorrected aberrated image (\textbf{D}).}
\label{fig:decon}
\end{figure}

The results are displayed in \figurename{\ref{fig:decon}} \textbf{A}, which shows SSIM as a function of p/px.
The p/px in \figurename{\ref{fig:decon}} \textbf{A} were multiplied by $4/3$ for the estimated corrected images to account for the extra photons needed to acquire the corrected image. The RWEs of the aberration estimations are also plotted, with values given on the right-hand scale.

It can be seen that both correction approaches substantially improve over the uncorrected image at high SNR. The DM-based re-acquisition provides a small but clear improvement over deconvolution at high SNR and nearly approaches the image quality provided by the hypothetical ideal correction. However, at lower SNR the extra photon dose combined with the poor estimation accuracy substantially decreases the quality of both correction approaches and leads to deconvolution outperforming the DM-corrected image, at least when dose matching as we have done here.

\section{Discussion}
The PD algorithms presented here estimate phase aberrations accurately enough for image improvement, even in the many detrimental settings tested in simulation. Unlike the use of SHWFS, PD requires no specialized hardware (even the DM is only required for correction), accounts for extended objects, and measures aberrations in the path to the image sensor. Consequently, PD algorithms can be fast, simple, and robust methods to measure phase aberrations. The PD algorithms can be run with uploaded images at \url{https://share.streamlit.io/nikolajreiser/PoissonPhaseDiversity}, and the code and usage instructions are available at \url{https://github.com/nikolajreiser/PoissonPhaseDiversity}.

The results here also show that aberration correction can be achieved with no specialized hardware at all. If defocus is used as a diversity phase by physically moving the lens, diversity images can be acquired without a DM, and the estimated aberrated PSFs then used in a multi-image deconvolution. At high SNR, the dose-matched performance of this strategy only slightly lags one that uses a DM to reacquire images corrected for the aberrations, and at low SNR it slightly outperformed the DM. While the process of wavefront estimation and deconvolution is technically equivalent to blind deconvolution, explicitly separating the wavefront estimation from deconvolution permits implementations tailored for each step.

While both PD algorithms achieve accurate phase estimations, the Poisson algorithm estimates the phase more accurately than the Gaussian algorithm in most cases. The Poisson algorithm is also less sensitive to effects not modeled in the LFs, such as spatial variance, phase noise, and out of focus plane contamination (see supplement for results on the latter).

There are many areas where further research could improve these algorithms. The optimization process could be improved for both algorithms, in terms of runtime and avoiding local minima. Also, diversity phases besides defocus were not explored. An improved method for diversity phase selection could enhance the algorithms, although it would likely depend strongly on specifics of the imaging system and the imaged object.

The PD algorithms should only be used for widefield detection. Imaging systems that are point scanning do not have to consider the object in phase retrieval, simplifying aberration estimation. Furthermore, the widefield PD algorithms are incapable of measuring spatial variation without subdividing the FOV. Smaller FOVs are likely to lead to poor aberration estimates, as seen in section \ref{imsize_section}, and re-acquisition would require multiple acquisitions with different DM settings tailored to the different FOVs.  

\section{Conclusion}
Poisson and Gaussian phase estimation algorithms were implemented and compared in the context of widefield microscopy. The Poisson algorithm is found to match or outperform the Gaussian algorithm for a variety of objects as well as for a range of illumination intensity and aberration levels. The Gaussian algorithm performs better in low-light regimes when image noise is dominated by additive Gaussian noise. The Poisson algorithm is also found to be more robust to the effects of spatially variant aberration and phase noise. Finally, the advantage of re-acquisition with aberration correction over deconvolution using the estimated aberrated PSFs is demonstrated. However, the performance of deconvolution is strong enough that it could  be used in situations where no deformable mirror is available. 

\begin{backmatter}
\bmsection{Funding}
This research is funded in part by the Gordon and Betty Moore Foundation. This research was also supported by the intramural research programmes of the National Institute of Biomedical Imaging and Bioengineering. Partial funding for this work was provided by the NIH S10-OD025081, S10-RR021039, and P30-CA14599 awards. The contents of this paper are solely the responsibility of the authors and do not necessarily represent the official views of any of the supporting organizations.

\bmsection{Disclosures}
The authors declare no conflicts of interest.

\bmsection{Data availability} 
Code and data underlying the results presented in this paper are available online at \url{https://github.com/nikolajreiser/PoissonPhaseDiversity}.

\bmsection{Supplemental document}
See Supplement 1 for supporting content. 

\end{backmatter}


\end{document}